\documentclass[fleqn,10pt]{wlscirep1}
\usepackage{changes}
\usepackage{lipsum}
\usepackage{multirow}
\usepackage{hyperref}
\usepackage{fix-cm}
\usepackage{amsmath}
\usepackage{graphicx}
\usepackage{verbatim}
\usepackage{tabularx}
\usepackage{ctable}
\usepackage{array}
\usepackage{xcolor}
\usepackage{colortbl}
\usepackage{hhline}
\usepackage{caption}
\usepackage{graphics}
\usepackage{epsfig}

\usepackage{etoolbox}
\makeatletter
\patchcmd{\@maketitle}{ABSTRACT}{}{}{}
\makeatother

\usepackage[linesnumbered,ruled,vlined]{algorithm2e}

\SetCommentSty{mycommfont}

\SetKwInput{KwInput}{Input}                
\SetKwInput{KwOutput}{Output}              

\usepackage{sidecap}

\title{Inference of the Dynamic Aging-related Biological Subnetwork via Network Propagation}

\author[1]{Khalique Newaz}
\author[1,*]{Tijana Milenkovi\'{c}}
\affil[]{Department of Computer Science and Engineering, Interdisciplinary Center for Network Science and Applications, and Eck institute for Global Health, University of Notre Dame, Notre Dame, IN 46556, USA}

\affil[*]{Corresponding author (email: tmilenko@nd.edu)}

\begin{abstract}
Gene expression (GE) data capture valuable condition-specific information (``condition'' can mean a biological process, disease stage, age, patient, etc.) However, GE analyses ignore physical interactions between gene products, i.e., proteins. Since proteins function by interacting with each other, and since biological networks (BNs) capture these interactions, BN analyses are promising. However, current BN data fail to capture condition-specific information. Recently, GE and BN data have been integrated using network propagation (NP) to infer condition-specific BNs. However, existing NP-based studies result in a static condition-specific subnetwork, even though cellular processes are dynamic. A dynamic process of our interest is human aging. We use prominent existing NP methods in a new task of inferring a dynamic rather than static condition-specific (aging-related) subnetwork. Then, we study evolution of network structure with age --  we  identify proteins whose network positions  significantly change with age and predict them as new aging-related candidates. We validate the predictions via e.g., functional enrichment analyses and literature search. Dynamic network inference via NP yields higher prediction quality than the only existing method for inferring a dynamic aging-related BN, which does not use NP. 
\end{abstract}
\begin{document}

\flushbottom
\maketitle

\thispagestyle{empty}

\section{Introduction}

\subsection{Motivation and related work}
Gene expression data, which have revolutionized our biomedical understanding \cite{cookson2009}, capture valuable condition-specific information (``condition'' can mean a biological process, disease stage, age, patient, etc.). However, gene expression analyses ignore connectivities between gene products (i.e., proteins) in a cell (we use terms ``gene'' and ``protein'' interchangeably). Yet, proteins interact to carry out cellular functions, and this is what protein-protein interaction (PPI) networks model. So, PPI network research can deepen our biomedical understanding \cite{caldera2017}. However, the current PPI network of a species spans many  conditions \cite{Yeger2015}. Using the PPI data alone without looking at other condition-specific (typically gene expression) data fails to capture  any condition-specific knowledge. 

Hence, recent studies integrated gene expression and PPI data via network propagation (NP), which maps gene activities (expression levels) onto the corresponding proteins in the PPI network. Then, NP propagates the activities via random walks or diffusion, to assign condition-specific weights to the nodes (i.e., proteins) or the edges (i.e., PPIs) in the network \cite{Cowen2017}. We note that besides NP  methods, there exist other, non-NP types of methods that integrate condition-specific data with PPI network data, such as kernel, Bayesian, or non-negative matrix factorization methods  \cite{mitra2013integrative,gligorijevic2015methods}. We focus on NP  methods, and so non-NP methods are out of the scope of our study. Also, we note that there exists another category of data integration approaches, which fuse condition-specific data with individual biological pathways as opposed to the whole PPI network \cite{li2013subpathway,garcia2015pathway,Vrahatis2016,koumakis2016,pietras2018tempo}. Because we are interested in the latter, the former approach category is out of the scope of our study.

Existing NP approaches can be grouped into two categories. One category are approaches for \emph{condition-specific node prioritization}. These approaches use NP to assign weights to nodes in the network, with a hypothesis that the higher the weight of a node, the more likely the node is to be related to the condition in question \cite{qian2014identifying,smedley2014walking}. Approaches of this type typically do not weigh edges. The other category, which is what we focus on in this paper, are approaches that focus on \emph{condition-specific subnetwork identification}. These approaches use NP to assign weights to edges (and sometimes also to nodes) in the network and then identify highly weighted network regions as a condition-specific subnetwork \cite{Mazza2016, Paul2013, Vandin2011}. Here, condition-specific gene information can be gene expression data, or it can be gene mutation data on e.g., how many patients have genes containing significantly associated single nucleotide polymorphisms, indels, etc. Two prominent methods from this category are NetWalk \cite{Komurov2010} and HotNet2 \cite{Leiserson2015}.

NetWalk integrates the condition-specific gene information with network topology \emph{immediately}, by  performing, from all nodes simultaneously, a random walk on the network  biased by the condition-specific gene information.  
On the other hand, HotNet2 first summarizes network topology in the form of a diffusion matrix, by performing, from one node at a time, an unbiased random walk on the network; this diffusion matrix captures the topological effect of each node on all other nodes in the network. \emph{Only then} HotNet2 combines the condition-specific gene information with the topology-based diffusion matrix. The new (final) diffusion matrix captures both the topological and condition-specific effect of each node on all other nodes.

Also, NetWalk and HotNet2 differ as follows. As its output, NetWalk assigns weights to all edges in the network. The edge-weighted network can then be used to identify a condition-specific subnetwork, by extracting only the highest-weighted network regions. However, as a part of its algorithm, NetWalk does not explicitly define a procedure for doing this, and one needs to devise it on their own. 
As its output, HotNet2 identifies a given group of nodes and all of their corresponding edges from the entire PPI network as a condition-specific subnetwork if the nodes in the group have strong mutual effects according to HotNet2's final diffusion matrix. Both methods have been used to study cancer, i.e., predict new cancer-related genes or molecular pathways \cite{Leiserson2015, ZHENG2016, bailey2016, Komurov2010}.

Another NP method exists that identifies each node's ``neighbor-network'', computes condition-specific activity of each neighbor-network via  its enrichment in highly expressed genes, and identifies all significantly active neighbor-networks as a single condition-specific subnetwork \cite{ansari2017approach}. We could not consider this approach in our study, because of the unavailability of the software at the time of our study.

Other types of NP-based subnetwork identification methods exist. NetQTL \cite{kim2011} and TieDIE \cite{paull2013} hypothesize that for a given condition, there is a set of source (e.g., transcription factor) genes that affect a set of target (e.g., differentially expressed) genes, and that the important network paths connecting the source genes to the target genes are a good representation of the condition-specific subnetwork. Hence, these methods propagate the condition-specific information from the source genes to the target genes, with the goal of identifying important paths between the source and target genes. Clearly, unlike NetWalk and HotNet2, NetQTL and TieDIE require two sets of condition-specific genes, i.e., sources and targets, in order to identify a condition-specific subnetwork. As such, NetQTL and TieDIE are out of the scope of our study.

The existing NP studies obtained a single, \emph{static} condition-specific PPI subnetwork. This is because they studied an $n\times 1$ vector containing gene expression/mutation information of $n$ genes for a single condition. Or, when they used an $n \times m$ matrix containing information of $n$ genes for $m$ conditions, they used  all $m$ conditions to compute a single activity value for the given gene, thus summarizing the matrix into an $n\times 1$ vector. For example, the authors of HotNet2 \cite{Leiserson2015} analyzed mutation weights of ${\sim}12,000$ ($n$) genes across ${\sim}3,000$ ($m$) samples related to different cancer types, which they then summarized into a  ${\sim}12,000\times 1$ vector, where a given position in the vector quantified the likelihood of the corresponding gene being active in many samples. Then, the summarized vector was used to obtain a single (not necessarily connected) condition-specific subnetwork, i.e., subnetwork active in several cancer types.

In contrast, cellular processes are \emph{dynamic}. This includes human aging, which we are interested in studying because the occurrence of diseases increases with age \cite{benz2008, plowden2004, rosenzweig2003}. Hence, studying aging, a dynamic process, via inference and analysis of a static aging-related subnetwork, can be limiting. Inferring and analyzing a dynamic aging-related subnetwork is expected to be more promising when the goal is to study temporal changes of network structure and thus cellular functioning with age. 

\subsection{Motivation}
Currently, there exists only one type of approach for inference of a dynamic aging-related subnetwork, which we refer to as the induced approach. Given gene expression data for multiple ages and a static PPI network, the induced approach identifies, for a given age, all proteins that are active (significantly expressed) at that age and all PPIs involving these proteins (i.e., it  extracts the induced subgraph among the active proteins). This results in a PPI subnetwork that is specific to the age in question. Repeating this for all ages results in one age-specific PPI subnetwork per age, which combined form a dynamic aging-related PPI subnetwork.

To the best of our knowledge, only two existing studies inferred a dynamic, aging-related PPI subnetwork, and both used the induced approach \cite{Faisal2014,elhesha2019identification}. Several additional studies used the aging-related subnetworks resulting from the induced approach in various computational tasks, such as  alignment of dynamic networks \cite{vijayan2018aligning, elhesha2019identification, aparicio2019temporal},   clustering of a dynamic network \cite{crawford2018cluenet, hulovatyy2016scout}, and studying changes in network positions of nodes (i.e., genes) with age \cite{Faisal2014, yoo2015improving}.
Moreover, several existing studies  inferred dynamic condition-specific PPI subnetworks relevant for studying  biological phenomena other than aging. These include a study of Prion disease \cite{newaz2015identification}, as well as studies aiming to identify protein complexes \cite{shen2017identifying, li2017identification,wang2014dynamic, chen2014identifying, zhang2016construction} or disease progression biomarkers \cite{wang2014dynamic}. Again, all of the existing studies have relied on the induced approach \cite{wang2014dynamic, chen2014identifying, zhang2016construction}, and they methodologically differ from each other mainly in how they defined a gene to be active at a given time point.

The induced approach considers \emph{all} interactions from the static network that exist between \emph{only active} genes. However, first, not all interactions between the active proteins might be equally ``important''. The induced approach has no mechanism of identifying only the most important (e.g., highly weighted) of all interactions between the active genes. NP methods for subnetwork identification, which are able to assign condition-specific weights to edges, can help. Second, it might be important to consider both active proteins and non-active proteins that critically connect the active proteins in the network, which the induced approach fails to do. NP can help, because it propagates activities of highly expressed nodes to other nodes in the network, thus possibly giving a high weight to a non-active node if e.g., it is surrounded by many active nodes or is on many paths between  active nodes. 

To address these drawbacks of the induced approach, we generalize the existing NP methods for static condition-specific subnetwork identification to their dynamic counterparts. We hypothesize that using NP rather than the induced approach will improve the quality of the inferred dynamic aging-related PPI subnetwork and thus yield higher-quality aging-related gene predictions. Note that we do not consider NP methods for node prioritization, because they have a different goal than our goal of condition-specific subnetwork identification.

\begin{figure*}[!h]
	\centering
\includegraphics[scale=0.97]{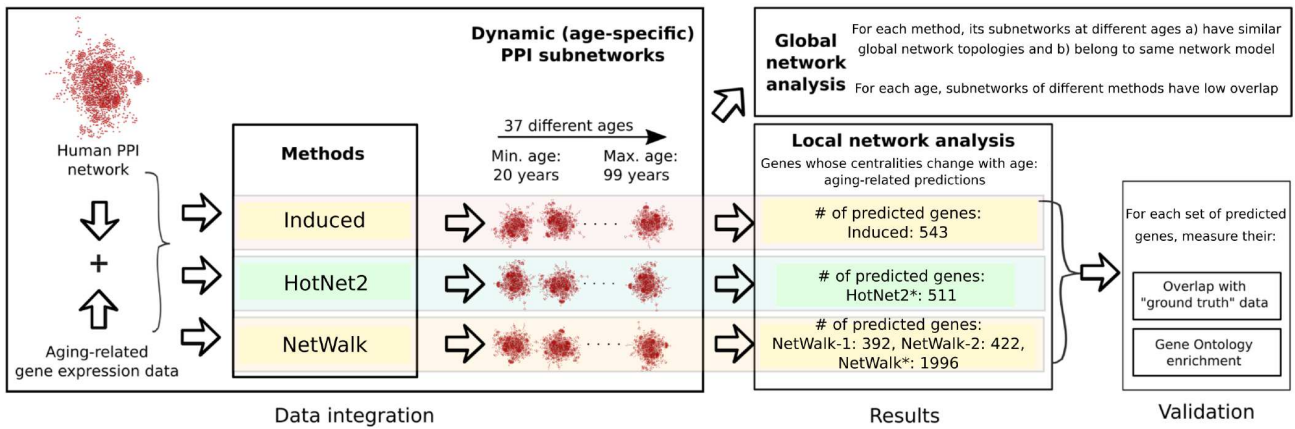}

    \caption{Summary of our study. We integrate a static human PPI network with gene expression data at different ages, using five versions of three approaches, which results in five dynamic aging-related PPI subnetworks. For each dynamic subnetwork, we study changes in global and local network topology with age. While global network topology does not change with age, we find significant changes in local network neighborhoods of 392 to 1,996 proteins, depending on the dynamic subnetwork. We predict such proteins as aging-related candidates. For the five dynamic subnetworks, this results in five aging-related gene prediction sets. We validate the prediction sets in several ways. For details on each step of our study, see Section \ref{sect:methods}.   
    }
    \label{fig:fig1}
\end{figure*}

\subsection{Our contributions}
We test our hypothesis using NetWalk and HotNet2  NP methods. While one might argue that other NP methods for subnetwork identification exist that could perhaps be used instead, showing that at least one of these two considered methods improves upon the induced approach is sufficient to confirm our hypothesis. Using any other potentially superior NP method would only further strengthen the superiority of NP over the induced approach. We use each of NetWalk, HotNet2, and the induced approach to integrate aging-related gene expression data with static human PPI network data, in order to construct a dynamic aging-related subnetwork corresponding to the given approach (Fig. \ref{fig:fig1}). 

After we study (dis)similarities between the dynamic subnetworks, we analyze which one is of the highest quality, i.e., the most relevant for the aging process. This can be done in either \emph{unsupervised} or \emph{supervised} manner. An unsupervised approach does not consider current knowledge about aging when making aging-related predictions but instead considers it only when evaluating the predictions, while a supervised approach considers a part of the
current knowledge about aging when making aging-related predictions and its other part when evaluating the predictions \cite{fabris2017review,li2019supervised}. In this study, we focus on the former; the latter is a complementary on-going work by our group \cite{li2019supervised}. Specifically, we use an established unsupervised framework for dynamic network analysis of aging that studies how network positions (network centrality values) of proteins change with age and predicts proteins that show significant changes with age as aging-related candidates \cite{Faisal2014}. We apply the framework to the dynamic subnetworks corresponding to the different approaches, resulting in a set of aging-related gene predictions for each approach. We validate, i.e., quantify the quality of, the predicted gene sets by measuring their overlaps with independent aging-related ``ground truth'' data and via functional enrichment analyses. 

We find that our NP-based predicted gene sets have  significant gene or functional overlaps with the ground truth data. Importantly, the overlaps are better for our NP-based predicted gene sets than for the gene set predicted by the induced approach. For example, GenAge, a trustworthy  aging-related ground truth dataset \cite{De2009}, shows an overlap of ${\sim}38\%$ with one of the NP approaches (adjusted $p$-value, i.e., $q$-value, of $1.3 \times 10^{-7}$), but an overlap of only ${\sim}8\%$ with the induced approach ($q$-value of $0.27$).

We find that all of the predicted gene sets, including that of the induced approach, contain novel aging-related predictions, i.e., genes that are not present in any of the ground truth datasets. Since the NP-based predicted gene sets show higher overlaps with the ground truth data than the induced approach's predicted gene set, we trust their novel predictions more than the novel predictions of the induced approach. Also, for each approach, we find some predictions that are unique to it, implying that the different approaches, i.e., the different dynamic subnetworks, are capturing at least somewhat complementary aging-related information. This is not surprising given that the node and edge overlaps between the dynamic subnetworks inferred by the different approaches are low.

We aim to link the novel NP-based predicted genes that are not predicted by the induced approach to aging via literature search. There are 16 such predictions, of which we validate nine, i.e., a majority of them.
 
\section{Methods}
\label{sect:methods}

\subsection{Static human PPI network data}
\label{sec:methods-ppidata}

We integrate a static PPI network with aging-related gene expression data to construct a dynamic aging-related PPI subnetwork. To study the robustness of our results to the choice of static PPI network, we perform our analyses on two different static PPI networks, as follows.

We use the human PPI network from HPRD \cite{Hprd2009}, the same data used in the dynamic network study of aging via the induced approach \cite{Faisal2014}. We extract the network's largest connected component, which has 8,938 nodes and 35,900 edges. We denote the set of 8,938 nodes as \emph{StatNetGenes}. We use the largest connected component and not the whole PPI network because the NP methods require a network to be connected. The HPRD PPI network is one of the networks of our interest because we aim to mimic the existing induced approach's study in all aspects except how the input static PPI network and aging-related gene expression data are integrated; that is,  we aim to use NP rather than the induced approach for data integration. The induced approach study focused in depth on the HPRD PPI data, which is why we use the same data here as well.

The induced approach study already showed that the choice of PPI data did not have a major effect on the quality of results (i.e., aging-related gene predictions). It did so by comparing its results obtained using the HPRD PPI network against its results obtained using the BioGRID PPI network \cite{stark2006}. Nonetheless, because in our study we use NP in addition to the induced approach and because the induced approach study did not consider NP at all, we again test the robustness of our results to the choice of static PPI network. Specifically, in addition to considering the HPRD PPI network, we use a recent HINT+HI2 PPI network data that combines two high-quality PPI datasets: the HINT PPI database that compiles high-quality PPI interactions from eight different PPI databases \cite{das2012hint} and the human HI2 PPI database \cite{rolland2014proteome}. We extract the network's largest connected component, which has 9,858 nodes and 40,704 edges. For informational purposes, 7,037 and 14,902 of these nodes and edges, respectively, are also in the HPRD network.

We run all of our following analyses on each of the HPRD and HINT+HI2 networks. For simplicity, we explain the following methodological analyses for the case of using the HPRD network. Additionally, since we find that results corresponding to the HPRD network are qualitatively similar to results corresponding to the HINT+HI2 network, for brevity, we report results for the HPRD network (henceforth referred to as the static PPI network). We discuss key results for the other network in Section \ref{sect:results-robust}.

\subsection{Aging-related gene expression data }
\label{data:geneexpr}
We use the same human aging-related gene expression data that was used in the induced approach study \cite{Faisal2014, Berchtold2008}. The  data encompasses 173 samples from 55 individuals' brains that span  37 different ages between 20 and 99 years. In order to identify whether a gene is significantly expressed (i.e., active) at a given age, we follow the procedure from the induced approach study: a gene is defined as \emph{active} if the $p$-value for its expression detection is below a threshold (Supplementary Section S1). Note that the induced approach study accounted for qualitative information (whether a gene is  active at a given age or not), but not for quantitative information about the actual gene expression values. The NP methods require as input the quantitative information, i.e., assigning scores to genes (see below). So, as gene scores, we use the gene expression values: an age-specific score for a gene is the average of the gene's log-scaled expression values across all samples for the given age. For details, see Supplementary Section S1.

We focus on this gene expression data rather than alternative data because we aim to  mimic the induced approach study in all aspects except how we integrate the input data. The induced approach study  focused \emph{in depth} on this data, which is why we use it here as well. Also, the induced approach  study \emph{briefly} analyzed an alternative gene expression data \cite{mazin2013widespread} and found the results to be qualitatively similar, i.e., the choice of aging-related gene expression data did not have a major effect.

\subsection{Integrating static PPI network with gene expression data to obtain dynamic aging-related PPI subnetworks }
\label{method:integration}
We use three methods to construct dynamic aging-related PPI subnetworks: the induced approach,  HotNet2, and  NetWalk. Additionally, for each of HotNet2 and NetWalk, we define two versions depending on how we assign age-specific scores to genes. 

\vspace{0.1cm}

\noindent\textit{The induced approach. } To construct a PPI subnetwork specific to a given age,  the induced approach selects all proteins that are active at that age and all PPIs among those proteins. The induced approach's collection of all 37 age-specific PPI subnetworks corresponding to the 37 considered ages forms a dynamic aging-related PPI subnetwork that we refer to as \emph{Induced}. 

\vspace{0.1cm}

\noindent\textit{HotNet2. }We run HotNet2 for each age as described below to get a PPI subnetwork specific to the given age. Then, HotNet2's collection of all 37 age-specific PPI subnetworks forms its dynamic aging-related PPI subnetwork.

\begin{figure*}[!t]
	\centering
	\includegraphics[scale=0.5]{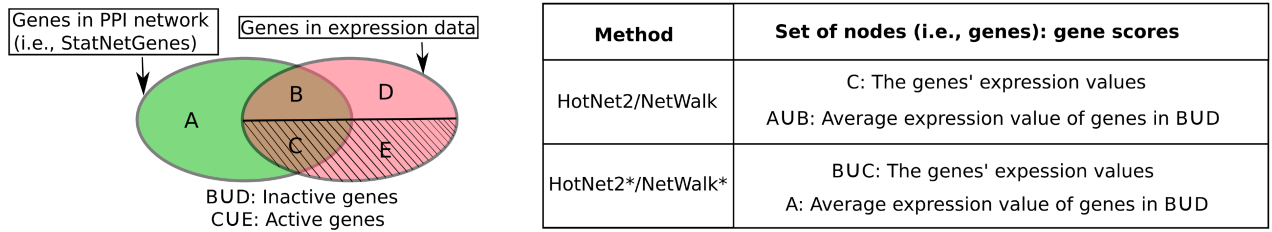}

    \caption{Illustration of the two different ways of assigning age-specific node scores to the nodes in the static PPI network. }
    \label{fig:node-score}
\end{figure*}

The HotNet2 study \cite{Leiserson2015} aimed to identify pancancer-related active PPI subnetwork. To do this, given pancancer mutation data of ${\sim}19,000$ genes across ${\sim}3,000$ samples, the HotNet2 study assigned  pancancer-related importance scores (i.e., mutation scores in the form of single nucleotide variations or copy number alterations) to those ${\sim}12,000$ genes that were sufficiently expressed (i.e., active). Finally, the HotNet2 study mapped the mutation scores of these active genes to their respective proteins in a PPI network and applied the HotNet2 algorithm to identify a single (not necessarily connected) pancancer-related active subnetwork.

To be able to use HotNet2 to integrate static PPI network data with age-specific gene expression data to obtain an age-specific subnetwork, we score genes using a similar strategy as that of the HotNet2 study. That is, we assign age-specific scores (see below) to only those StatNetGenes genes that are \emph{active} according to the gene expression data (see Section \ref{data:geneexpr} for our definition of an active gene). Given that we consider gene expression data as opposed to gene mutation data that the HotNet2 study considered, we use the active genes' age-specific expression values (Section \ref{data:geneexpr}) as their age-specific scores, as opposed to the mutation scores that the HotNet2 study considered. For each of the StatNetGenes genes that are \emph{not active} according to the gene expression data or are entirely missing from the gene expression data, we use as their scores the average gene expression value of all non-active genes from the gene expression data. We do this because if we did not, i.e., if instead we ignored these StatNetGenes genes by giving them a score of zero, then HotNet2 would not consider them at all in the construction of an age-specific subnetwork. We want to avoid this, because it would not allow us to test our hypothesis that proteins that are not actively present in the gene expression data but that critically connect active proteins should be a part of the condition-specific subnetwork. We identify this version of HotNet2 as \emph{HotNet2}.

Note that the definition of whether a gene is active or not depends on the protocol used to identify active genes in a gene expression data, including but not limited to the $p$-value cutoff for expression detection \cite{park2003,mcclintick2006}. Hence, depending on the $p$-value cutoff for example, some genes that are actually important in the gene expression data can be identified as unimportant, which could bias results. In order to avoid this, we define another version of HotNet2, \emph{HotNet2*}. In this version, instead of using gene expression values as scores for only the active StatNetGenes genes but using the average gene expression value of all non-active genes as scores for all other StatNetGenes (i.e., for non-active StatNetGenes genes and those genes absent from the gene expression data), we use gene expression values as scores for \emph{all} of those StatNetGenes genes that are present in the gene expression data, regardless of whether they are active or not; for the remaining StatNetGenes gene, i.e., those absent from the gene expression data, we use as their scores the average gene expression value of all non-active genes. 

For each of the two ways of assigning age-specific scores to genes (i.e., HotNet2 and HotNet2*), which are summarized in Fig. \ref{fig:node-score}, we apply the HotNet2 algorithm to obtain the corresponding age-specific subnetwork. Specifically, the HotNet2 algorithm first performs network diffusion to combine gene scores with the network topology. This results in a diffusion matrix $M$ where intuitively each value $M_{ij}$ quantifies the effect of node $i$ on node $j$. The matrix $M$ contains an entry for each node pair $i$ and $j$ in the network, independent of whether the two nodes are linked by an edge in the network or not. The HotNet2 algorithm keeps only those matrix values that are greater than or equal to a given threshold and converts all $M_{ij}$ values below the threshold to zero. Then, it uses this updated matrix to: 1) create a directed network such that there is a directed edge from node $i$ to node $j$ if and only if the corresponding value in the updated matrix is not zero; 2) identify strongly connected components in the directed network; 3) take the union of nodes present in all of the identified strongly connected components; 4) map these nodes to the static PPI network; and 5) extract an induced PPI subgraph on the nodes as the final output, i.e., a condition-specific PPI subnetwork. 

To select a threshold value from matrix $M$, the HotNet2 algorithm requires the user to specify the desired minimum size of the strongly connected components of the output subnetwork, as well as the use of a random network model. However, there is no prior ground truth knowledge about what the size of an age-specific PPI subnetwork should be. Also, using different random network models can give different results \cite{yaverouglu2014revealing}. 

To avoid these problems, we examine 100 different threshold values and generate 100 dynamic aging-related subnetworks, as discussed below. 

We vary the threshold as follows. 
Ideally, we would have divided all of the matrix $M$ values  into 100 equal-size bins, and used some (e.g., the minimum) value in the given bin as a representative threshold. However, we  found that the HotNet2 algorithm's matrix $M$ values follow an unusual distribution in the sense that considering as a threshold any value that is not among the top 0.1\% highest values would return as the output (i.e., as an age-specific subnetwork) almost the entire, if not the entire, static PPI network. This would not be useful, because we aim to test whether removing less important regions of the entire network improves quality of predictions made from the network. That is, the most interesting threshold choices lie only among the top 0.1\% highest matrix values. This is why we only consider these values, by dividing them into 100 equal-size bins and using the minimum value in each bin as one of the 100 thresholds. 

This way, we obtain 100 different dynamic subnetworks corresponding to HotNet2 and 100 different dynamic subnetworks corresponding to HotNet2*, which we use as discussed in Section \ref{sec:methods-select}. Algorithm \ref{algo:1} outlines the above procedure of obtaining 100 dynamic subnetworks with a given approach.

\vspace{0.1cm}
\begin{algorithm}[H]
\DontPrintSemicolon
  Let $DAS$ be an empty ordered set. \tcp{To store dynamic  subnetworks}

  Let $AS$ be an empty 2-dimensional matrix. \tcp{To store all age-specific subnetworks}

  \vspace{0.1cm}
   \For{$i^{th}$ age from 1 to 37}    
        { 
        	Use gene expression values corresponding to $i^{th}$ age to score nodes in $G$ to obtain a node-scored network ($G_{i}^{ns}$) \tcp{Fig. 2}
        	Apply algorithm $P$ on $G_{i}^{ns}$ to obtain an edge-weighted network ($G_{i}^{ew}$)
        	Determine 100 edge weights among edge weights in $G_{i}^{ew}$ as 100 thresholds \tcp{Section 2.3}
      	
        	\For{$j^{th}$ threshold from 1 to 100}    
        {
            Let $AS_j$ be an empty set \tcp{To store the age-specific subnetwork for the $j^{th}$ threshold}
            
            \For{each edge $e$ in $G_{i}^{ew}$}
            {
                \If{weight($e$) $\geq$ the $j^{th}$ threshold}
                {
                $AS_j$ = UNION($AS_j$,$e$) \tcp{Add the edge }
                }
            }
            
            $AS[i][j]$ = $AS_j$ \tcp{Store the age-specific subnetwork for the $j^{th}$ threshold}
 
        }
        
        }
        
    \For{$j^{th}$ threshold from 1 to 100}
    {
        Let $DAS_j$ be an empty ordered set \tcp{To store the dynamic subnetwork for the $j^{th}$ threshold}
    
    \For{$i^{th}$ age from 1 to 37}
    {
        $DAS_j[i]$ = $AS[i][j]$ \tcp{Add the subnetwork for the $j^{th}$ threshold of the $i^{th}$ age}
    }
    
    $DAS[j]$ = $DAS_j$ \tcp{Store the dynamic subnetwork for the $j^{th}$ threshold}

}
\Return{$DAS$}
\caption{Given a static PPI network $G$, a network propagation algorithm $P$, and all genes' expressions for $i^{th}$ age (where, $i$ = 1, 2, ..., 37), return a dynamic subnetwork for each threshold $j$ (where, $j$ = 1, 2, ..., 100).
\label{algo:1}
}
\end{algorithm}
\vspace{0.1cm}


\vspace{0.1cm}
\noindent\textit{NetWalk. }Just as with the HotNet2 algorithm, we run the NetWalk algorithm for each age to get a PPI subnetwork specific to the given age. Then, NetWalk algorithm's collection of the 37 age-specific PPI subnetworks forms its dynamic aging-related subnetwork.

We assign age-specific scores to genes in the same two ways as we have done it for HotNet2 above, resulting in two versions of NetWalk: NetWalk and NetWalk* (these are analogs of HotNet2 and HotNet2*, respectively, as shown in Fig. \ref{fig:node-score}).

For each of the two ways of assigning age-specific scores to genes (i.e., NetWalk and NetWalk*),  we  apply the NetWalk  algorithm  to  obtain  the corresponding age-specific subnetwork. The output of the NetWalk algorithm is the entire input network but with each edge ($i,j$) being assigned two age-specific weights, one from node $i$ to node $j$, and the other one from  $j$ to  $i$. The NetWalk algorithm does not provide a procedure for extracting a condition-specific subnetwork from the  weighted output network. So, we need to design such a procedure, and we do it as follows.

Given the weighted network with bi-directional edge weights, for each edge, we select the minimum of its two bi-directional edge weights as the final edge weight. This way, an edge will be included into a subnetwork if and only if both of its bi-directional edge weights are larger or equal to a given threshold. Then, we divide all of the resulting edge weights into 100 equal-size bins. In each bin, we take the minimum of its edge weight values as a threshold. Then, for each of the 100 thresholds, we keep only those edges whose weights are equal to or greater than the given threshold.

This way (Algorithm \ref{algo:1}), we obtain  100  different  dynamic  subnetworks  corresponding to  NetWalk and  100  different dynamic  subnetworks  corresponding to NetWalk*, which we use as discussed in Section \ref{sec:methods-select}.

\vspace{0.1cm}
 \noindent\emph{Note.} Only for HotNet2, HotNet2*, NetWalk, and NetWalk* approaches, we can vary their respective thresholds to produce multiple (in our case, 100) different dynamic subnetworks. This can not be done for the Induced approach, as this approach does not rely on a threshold or any other parameter that allows for this. 

\subsection{Aging-related ground truth data }
\label{sect:methods-gtdata}
We use three sets of highly trustworthy aging-related ground truth data, possibly the best ones that are currently available.

We use a set of 305 human genes from GenAge that have been implicated in aging mostly because their sequence orthologs in model species have been shown to be aging-related \cite{De2009}. Of these 305 genes, 276 are present in StatNetGenes. We denote the 276 genes as \emph{GenAge}.

We use two other sets of genes from a recent study that analyzed aging-related genes derived from the Genotype Tissue-Expression project (GTEx) \cite{jia2018}. One of the gene sets contains 863 genes that have been shown to be down-regulated (i.e., their expressions decrease) with age. Of these 863 genes, 469 genes are present in StatNetGenes. We denote the 469 genes as \emph{GTEx-down}. The other gene set contains 710 genes that have been shown to be up-regulated (i.e., their expressions increase) with age. Of these 710 genes, 374 genes are present in StatNetGenes. We denote the 374 genes as \emph{GTEx-up}. It has been shown that the two sets of aging-related genes, i.e., GTEx-down and GTEx-up, show very different characteristics. Namely, GTEx-down genes are more likely to be evolutionary conserved, are significantly enriched in essential genes, are critical for PPIs, and show gene expression patterns that are not tissue-specific, while GTEx-up genes are less likely to be evolutionary conserved, are comparatively less enriched in essential genes, are not critical for PPIs, and show tissue-specific gene expression patterns \cite{jia2018}. Because GTEX-down genes are critical for PPIs while GTEx-up genes are not, and because we use PPI network data to predict aging-related genes,
we  expect  our  predicted gene sets  to  overlap more with GTEx-down than with GTEx-up.

\subsection{(Dis)similarities between dynamic subnetworks of different approaches}
\label{methods:global}

We evaluate (dis)similarities between any two dynamic subnetworks by: (1) measuring their pairwise node and edge overlaps, (2) comparing their global network properties (average clustering coefficient, average diameter, and graphlet degree distribution), to see whether the networks have (dis)similar topologies, and (3) evaluating their fit to five graph families or network models (Erdos-Renyi random graphs (ER), generalized ER with same degree distribution as the data network (ER-DD), geometric random graphs (GEO), scale-free network model (SF), and sticky graph model (Sticky), to see whether the networks belong to the same or different graph families. For details, see Supplementary Section S2.
 
\subsection{Do global topologies of a dynamic subnetwork change with age? }
For each dynamic subnetwork, we evaluate whether its global topology changes with age. Namely, we: (1) measure  pairwise node and edge overlaps between its age-specific subnetworks, (2) compare its age-specific subnetworks with respect to the three global network properties (see above), and (3) evaluate the fit of each age-specific subnetwork to the five network models (see above). For details, see Supplementary Section S2.

\subsection{Do local topologies of proteins in a dynamic subnetwork change with age? }
\label{methods:local}

For each dynamic subnetwork, we study topological positions of nodes (i.e., proteins) in each of its 37 age-specific  subnetworks using six network centrality measures: degree centrality (Degc), clustering coefficient centrality (Clusc), k-coreness centrality (Kc), graphlet degree centrality (Gdc), closeness centrality (Closec), and eccentricity centrality (Ecc) \cite{Faisal2014}, in order to predict as aging-related those genes whose centrality values significantly change with age. To do this, we rely on an established computational framework for dynamic network study of aging, which was proposed along with the induced approach \cite{Faisal2014}. Given a dynamic subnetwork, the framework computes, for each centrality measure, the centrality value of each node in each of the 37 age-specific PPI subnetworks. Then, it computes the Pearson correlation between the given node's centrality values and the 37 ages, and the statistical significance (i.e., $p$-value) of this correlation, where the $p$-value is the percentage of 1,000,000 random runs in which the random correlation is better than the actual correlation, which is then adjusted using the Benjamini-Hochberg procedure \cite{benj1995} to account for multiple testing correction. The framework predicts a gene as aging-related if for \emph{at least one} centrality measure, the gene's $q$-value is $< 0.01$, i.e., if its network position (centrality) significantly changes (increases or decreases) with age.

\vspace{0.15cm}

\subsection{Validation of predicted aging-related genes}
\label{method:validation}
We validate a given set of predicted aging-related genes in the following ways.

First, we believe that a predicted gene set is relevant for aging if it contains a significant number of genes that are already known to be aging-related according to independent knowledge that has not been used to make the predictions. Hence, we examine whether the predicted gene set significantly overlaps with an independent aging-related ground truth gene set (Section \ref{method:overlap}).

Second, we believe that a predicted gene set is relevant for aging if the genes in the set are involved in aging-related biological functions. Hence, we examine whether there is significant overlap between functions of the predicted genes and functions of aging-related ground truth genes. Specifically, given a gene set (i.e., a set of predicted genes or a set of aging-related ground truth genes), we identify all Gene Ontology (GO) terms that are significantly enriched (i.e., that annotate a statistically significant number of genes) in the gene set (Section \ref{method:go}). Then, given all GO terms enriched in a set of predicted genes and all GO terms enriched in a set of aging-related ground truth genes, we compute statistical significance of the overlap of the two sets of GO terms (Section \ref{method:goverlap}).

Third, we link the predicted genes to aging via literature search (Section \ref{method:litval}).

\subsubsection{Overlap between gene sets}
\label{method:overlap}
As is typically done, we use the hypergeometric test \cite{falcon2008} to compute the probability (i.e., $p$-value) of obtaining by chance the observed or higher overlap between two gene sets. Formally, if $S$ is StatNetGenes, $A$ is one of the two sets, $B$ is the other set, and $O$ is the overlap between $A$ and $B$, then the $p$-value is computed as: 

\begin{equation}
\label{eq3}
P(X \geq |O|) = 1 - \sum_{i=0}^{|O|-1} \frac{\binom{|S|}{i}\binom{|S|-|A|}{|B|-i}}{\binom{|S|}{|B|}}
\end{equation}

We compute overlaps for multiple pairs of gene sets, resulting in multiple $p$-values. To account for multiple hypothesis correction, we adjust the multiple $p$-values using the Benjamini-Hochberg procedure \cite{benj1995}, and  obtain the corresponding  $q$-values. We say that two gene sets overlap statistically significantly if the corresponding $q$-value is $\leq 0.01$.

\subsubsection{Gene Ontology (GO) term enrichment in a gene set}
\label{method:go}
We study enrichment of a gene set in a GO term by examining whether the GO term annotates a statistically significant number of genes in the gene set. We use all of those 9,464 GO terms that annotate at least two genes in StatNetGenes \cite{g2018}. As is typically done, we use the hypergeometric test \cite{falcon2008} to compute the likelihood of obtaining the given enrichment by chance. Intuitively, given the observed number of occurrences of a GO term in a gene set of the given size, the hypergeometric test measures the probability of getting the same or higher number of occurrences of the same GO term in a randomly chosen gene set of the same size. The latter is selected by chance from a set of background genes; as this set, we use all of those StatNetGenes genes that are annotated by at least one of the above mentioned 9,464 GO terms.
Given multiple enrichment $p$-values, one $p$-value for each of the GO terms that annotates at least one gene in the given gene set, we adjust the $p$-values into $q$-values as above, i.e., by using the Benjamini-Hochberg procedure. We say that a GO term is statistically significantly enriched in the gene set if its enrichment $q$-value is $\leq 0.01$.

\subsubsection{Overlap between enriched GO term sets}
\label{method:goverlap}
As is typically done, we use the hypergeometric test \cite{falcon2008} to compute the likelihood of obtaining by chance the observed or higher overlap between two sets of enriched GO terms (equation \ref{eq3}). Formally, in equation \ref{eq3}, $S$ is now the set of GO terms that annotate at least two genes from StatNetGenes, $A$ is the set of GO terms enriched in one of the datasets, $B$ is the set of GO terms enriched in the other dataset, and $O$ is the overlap between $A$ and $B$. 
We compute overlaps for multiple pairs of enriched GO term sets, resulting in multiple $p$-values. We adjust the $p$-values into  $q$-values as above, i.e., by using the Benjamini-Hochberg procedure. We say that two GO term sets overlap statistically significantly if the corresponding $q$-value is $\leq 0.01$.

\subsubsection{Literature validation} 
\label{method:litval}
We aim to link predicted genes to aging by searching for and closely reading relevant research articles in PubMed (\url{https://pubmed.gov}) or Google Scholar (\url{https://scholar.google.com}), using two key search phrases: 1) official symbol of a gene and ``aging'' and 2) official symbol of a gene and ``Alzheimer's''.

\subsection{Selection of a representative dynamic subnetwork for each of HotNet2, HotNet2*, NetWalk, and NetWalk*}
\label{sec:methods-select}
Recall that for each of HotNet2, HotNet2*, NetWalk, and NetWalk*, we obtain 100 dynamic subnetworks (Section \ref{method:integration}). To test our hypothesis that NP can improve upon the induced approach, we need to compare NP-based dynamic subnetwork(s) with the dynamic subnetwork of the induced approach. Hence, for each of HotNet2, HotNet2*, NetWalk, and NetWalk*, we ideally wish to select one of their respective 100 dynamic subnetworks that is the most relevant for aging, as follows.

Given 100 dynamic subnetworks, first, we use each of them to predict its respective aging-related genes via the framework from Section \ref{methods:local}. Second, to examine which of the 100 resulting  predicted gene sets are relevant for aging, we measure the statistical significance (i.e., $q$-value) of overlap of each of them with each of the aging-related ground truth datasets (Section \ref{sect:methods-gtdata}), using the procedure from Section \ref{method:overlap}. Then, for each ground truth dataset, we select the dynamic subnetwork whose prediction set shows the most significant (i.e., lowest) $q$-value.  If more than one prediction set shows such a $q$-value, i.e., if there is a tie, then we select the prediction set whose corresponding dynamic subnetwork has the fewest nodes, to keep as small number of genes in the subnetwork as possible, while capturing as complete  aging-related information as possible. If none of the prediction sets show a significant overlap, then we do not choose any dynamic subnetwork. Thus, we obtain \emph{at most} one dynamic subnetwork for each approach and each ground truth dataset combination. The results are:

\begin{itemize}

\item For HotNet2, the above procedure does not result in any dynamic subnetwork.

\item For HotNet2*, the above procedure results in a dynamic subnetwork corresponding to GenAge, which we refer to as HotNet2*.

\item For NetWalk, the above procedure results in two dynamic subnetworks: one corresponding to GenAge, which we refer to as NetWalk-1, and another corresponding to GTEx-down, which we refer to as NetWalk-2.

\item For NetWalk*, the above procedure results in the same dynamic subnetwork corresponding to both GenAge and GTEx-down, which we refer to as NetWalk*.

\item In total, over all NP approaches, we obtain the four NP-based dynamic subnetworks: HotNet2*, NetWalk-1, NetWalk-2, and NetWalk*.

\end{itemize}

\section{Results and discussion}

We integrate aging-related gene expression data with static human PPI network data using different approaches (HotNet2, NetWalk, and the induced approach) to obtain five aging-related dynamic subnetworks (HotNet2*, NetWalk-1, NetWalk-2, NetWalk*, and Induced). Given these dynamic aging-related subnetworks, we first study and compare their global network topologies (Section \ref{results:globalcomparison1}). Second, for each dynamic subnetwork, we study changes in its global network topology with age (Section \ref{results:globalcomparison3}). Third, in order to predict aging-related genes, for each dynamic subnetwork, we study changes in local topologies of its proteins with age, and predict the significantly-changing nodes as aging-related (Section \ref{results:localcomparison}). Given the aging-related gene predictions by the different dynamic subnetworks, we evaluate whether the NP-based dynamic subnetworks produce higher-quality predictions than the Induced dynamic subnetwork, which is exactly what we observe (Section \ref{results:validation}). As a negative control, we argue that randomized versions of the NP-based dynamic subnetworks should result in fewer predicted genes than the  actual dynamic subnetworks, and this is exactly what we observe (Section \ref{sec:validate-net}). Because genes present in multiple prediction sets are more likely to be aging-related, we identify genes present in all four NP-based prediction sets (resulting from the four NP-based dynamic subnetworks) but absent from every considered aging-related ground truth dataset as our novel gene predictions, and we aim to validate them using literature search (Section \ref{sect-results-litval}). Finally, we demonstrate that our key result -- the aging-related predictions made from the NP-based aging-related dynamic subnetworks being of higher quality than the aging-related predictions made from the Induced dynamic subnetwork -- is robust to the choice of static PPI network data (Section \ref{sect:results-robust}).

\subsection{Dynamic subnetworks contain different nodes and edges but show similar global topologies}
\label{results:globalcomparison1}
Since we use different data integration methods to create the dynamic aging-related subnetworks, we would not be surprised if the networks are at least somewhat dissimilar. We can expect differences between Induced and any of the four NP-based dynamic subnetworks (NetWalk-1, NetWalk-2,  NetWalk*, and HotNet2*), because of the different algorithmic mechanisms behind their  underlying data integration approaches. Also, we can expect differences between NetWalk-1, NetWalk-2, or NetWalk* on one side and HotNet2* on the other, because the former three use a different NP algorithm compared to the latter one.  Also, we can expect differences between NetWalk-1 or NetWalk-2 on one side and NetWalk* on the other, because the former two use a different  method to account for age-specific genes scores compared to the latter one (Section \ref{method:integration}). At the same time, because NetWalk-1, NetWalk-2 and NetWalk* share the underlying NP algorithm, we can expect some similarities between these networks.

\begin{figure*}[!t]
	\centering
	\includegraphics[scale=0.7]{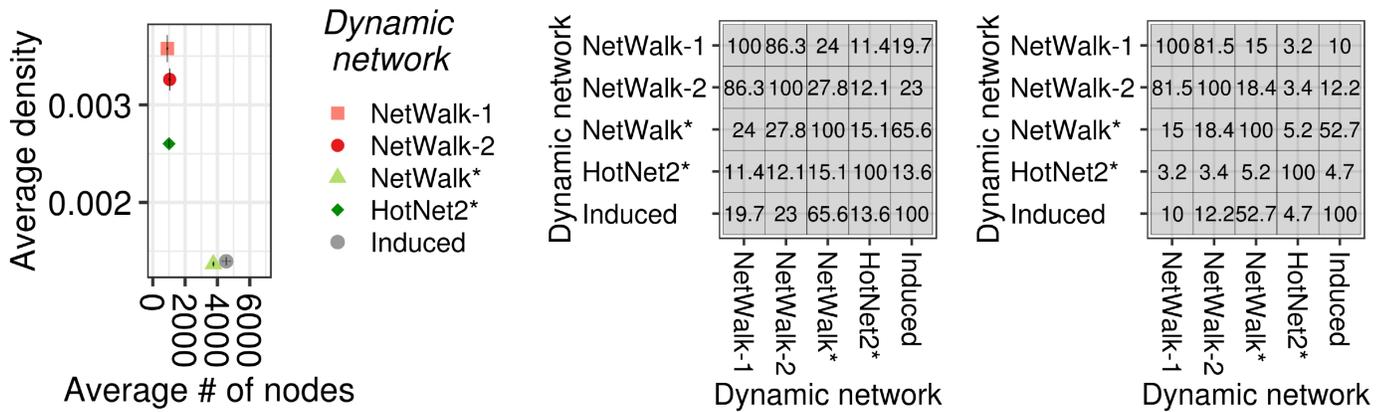}

    \caption{Comparison of the five dynamic subnetworks in terms of node size and density (left), node overlaps (middle), and edge overlaps (right). In the left panel, each point is the average over the 37 considered ages, and the vertical and  horizontal lines (if visible) represent the corresponding standard deviations. In the middle and right panels, each table cell shows the average overlap over the 37 ages.}
    \label{fig:dens}
\end{figure*}

First, we discuss similarity in terms of  sizes (numbers of nodes and densities) of the five dynamic subnetworks. 
\emph{In terms of the number of nodes}, three of the subnetworks, HotNet2*,  NetWalk-1, and NetWalk-2, all have $\sim$1,500 nodes, while the other two subnetworks, NetWalk* and Induced, have $\sim$4,000 nodes (Fig. \ref{fig:dens} (left) and Supplementary Fig. S1). Thus, surprisingly, Induced is more size-similar to an NP-based subnetwork than the NP-based subnetworks are to each other. Also, surprisingly,  NetWalk-1 and NetWalk-2 are more similar to HotNet2* than they are to NetWalk*. \emph{In terms of density}, while HotNet2*,  NetWalk-1, and NetWalk-2 have a similar number of nodes, they have different densities, i.e., some of these subnetworks have more edges than others. On the other hand, NetWalk* and Induced have not only a similar number of nodes but also a similar density. 

Second, we discuss similarity in terms of node/edge overlaps between the dynamic subnetworks (Supplementary Figs. S2 and S3). First, we focus on overlaps between Induced and the four NP-based dynamic subnetworks. Second, we focus on overlaps among the NP-based dynamic subnetworks themselves.

When analyzing overlaps between Induced and each NP-based dynamic subnetwork, \emph{in terms of node overlaps}, we find that Induced has a relatively high (${\sim}66\%$) overlap with NetWalk* but relatively low (${\sim}14\%$ to ${\sim}23\%$) overlaps with NetWalk-1, NetWalk-2, and HotNet2* (Fig. \ref{fig:dens} (middle)).  \emph{In terms of edge overlaps}, we find that Induced has a relatively high (${\sim}53\%$) overlap with NetWalk* but relatively low (${\sim}5\%$ to ${\sim}12\%$) overlaps with NetWalk-1, NetWalk-2, and HotNet2* (Fig. \ref{fig:dens} (right)). For NetWalk-1 and NetWalk-2, it is only the case that Induced contains many additional nodes/edges compared to NetWalk-1 and NetWalk-2 but not vice versa, while for NetWalk* and HotNet2*, it is both the case that Induced contains many additional nodes/edges compared to NetWalk* and HotNet2* as well as that NetWalk* and HotNet2* contain many additional nodes/edges compared to Induced (Supplementary Tables S1 and S2). 

When we analyze overlaps between the four NP-based dynamic subnetworks themselves, we find node overlaps to range from ${\sim}$11\% to ${\sim}$86\% (Fig. \ref{fig:dens} (middle)). As expected, NetWalk-1 shows the highest overlap with NetWalk-2, and both NetWalk-1 and NetWalk-2 show higher  overlaps with NetWalk* than with HotNet2*. The results for edge overlaps are similar (Fig. \ref{fig:dens} (right)).

Third, we discuss similarity in global  properties of the five dynamic subnetworks. We find that their average clustering coefficients, average diameter, and graphlet degree distributions  are similar. An exception is that Induced  shows a relatively higher average clustering coefficient compared to the NP subnetworks, and the other exception is that HotNet2*  shows a relatively  higher average diameter compared to the other four subnetworks (Supplementary Figs. S4 and S5). The best-fitting network models are the same in almost all cases subnetworks. Namely, both GEO and  Sticky are the best-fitting models for all dynamic subnetworks except for HotNet2*, where only GEO is the best (Supplementary Fig. S6).

\subsection{Global network topologies do not change with age}
\label{results:globalcomparison3}

In the previous section, we compared the five dynamic subnetworks to each other. Here, for a given dynamic subnetwork, we compare it to itself at different ages, i.e., we compare its 37 age-specific subnetworks to each other, to see whether its global network topology changes with age. We observe the following. First, for all of Induced, Netwalk*, and HotNet*, i.e., for a majority of the considered networks, pairwise node and edge overlaps of the age-specific subnetworks are relatively large (most are over ${\sim}$75\% ). For NetWalk-1 and NetWalk-2,  pairwise node and edge overlaps are somewhat lower (most are below 60\%); Supplementary Figs. S7 and S8. Second, for all of the dynamic subnetworks, the average clustering coefficients, average diameters, and graphlet degree distributions of the age-specific subnetworks are overall stable with age (Supplementary Figs. S4 and S5). Third, for all of the dynamic subnetworks, the age-specific subnetworks belong to the same network model(s) (Supplementary Fig. S6). So, we conclude that overall the global network topologies do not change with age.

\subsection{Local topologies of some proteins change with age, which are predicted as aging-related candidates}
\label{results:localcomparison}



Mimicking the induced approach study \cite{Faisal2014}, for each of the five dynamic subnetworks (NetWalk-1, NetWalk-2, NetWalk*, HotNet2*, and Induced), we use network centrality measures to study how local topologies of proteins change with age, and to predict as aging-related those proteins whose network centrality values significantly correlate with age (Section \ref{methods:local}). 

This results in  five predicted aging-related gene sets corresponding to the five dynamic subnetworks.  For simplicity, we denote the gene sets just as the corresponding dynamic subnetworks, i.e., NetWalk-1, NetWalk-2, NetWalk*, HotNet2*, and Induced. The number of predicted genes varies from ${\sim}4\%$ (NetWalk-1) to ${\sim}22\%$ (NetWalk*) of all 8,938 proteins in the static PPI network (Table \ref{tab:tab1} and Supplementary Figs. S9 and S10). 

\subsection{The five predicted gene sets significantly  overlap}  

Since we use the same input data to obtain all five predicted aging-related gene sets, we expect a significant overlap between them. Indeed, we find that while the overlaps are far from perfect, they are statistically significantly high for all pairs of predicted gene sets, except for the overlap between Induced and HotNet2* prediction sets, which is still marginally significant (Fig. \ref{fig:dens} and Table \ref{tab:tab0}). 

\begin{table}[ht]
\scriptsize
\centering
\setlength\tabcolsep{2pt}
\caption{Pairwise gene overlaps between the predicted gene sets. For the given set, its size is in parentheses. For each of the gene set pairs, i.e., in each table cell, we show (from the top): 1) the number of genes in the overlap, 2) overlap size as the percentage of the size of the union of the two sets, 3)  overlap size as the percentage of the size of the smaller of the two sets, and 4) the $q$-value of the overlap. Significant $q$-values (i.e., $q$-values $\leq 0.01$) are bolded.} 
\label{tab:tab0}
\begin{tabular}{|p{1.2cm}|p{1.2cm}|p{1.2cm}|p{1.1cm}|p{1.1cm}|p{1cm}|p{1.2cm}|}
  \hline
 & NetWalk-1 \newline (396)& NetWalk-2 \newline (422)& NetWalk* \newline (1996)& HotNet2* \newline (511)& Induced \newline (543)\\ 
  \hline
NetWalk-1 \newline (392) &  392 \newline 100\% \newline 100\% \newline \textbf{1e-300} &  311 \newline 61.8\% \newline 79.3\% \newline \textbf{1e-300} &  285 \newline 13.6\% \newline 72.7\% \newline \textbf{1.4e-105} & 52 \newline 6.11\% \newline 13.3\% \newline \textbf{7.7e-09} &  59 \newline 6.74\% \newline 15.1\% \newline \textbf{6.0e-11} \\ 
  \hline NetWalk-2 \newline (422) &  311 \newline 61.8\% \newline 79.3\% \newline \textbf{1e-300} &  422 \newline 100\% \newline 100\% \newline \textbf{1e-300} &  306 \newline 14.5\% \newline 72.5\% \newline \textbf{4.5e-113} &  54 \newline 6.14\% \newline 12.8\% \newline \textbf{1.2e-08} &  60 \newline 6.63\% \newline 14.2\% \newline \textbf{3.7e-10} \\ 
  \hline NetWalk* \newline (1996)&  285 \newline 13.6\% \newline 72.7\% \newline \textbf{1.4e-105} &  306 \newline 14.5\% \newline 72.5\% \newline \textbf{4.5e-113} &  1996 \newline 100\% \newline 100\% \newline \textbf{1e-300} &  189 \newline 8.15\% \newline 37\% \newline \textbf{7.0e-15} &  266 \newline 11.7\% \newline 49\% \newline \textbf{1.5e-45} \\ 
  \hline HotNet2* \newline (511)&  52 \newline 6.11\% \newline 13.3\% \newline \textbf{7.7e-09} &  54 \newline 6.14\% \newline 12.8\% \newline \textbf{1.2e-08} &  189 \newline 8.15\% \newline 37\% \newline \textbf{7.0e-15} &  511 \newline 100\% \newline 100\% \newline \textbf{1e-300} &  43 \newline 4.25\% \newline 8.41\% \newline 0.016 \\ 
  \hline Induced \newline (543) & 59 \newline 6.74\% \newline 15.1\% \newline \textbf{6.0e-11} &  60 \newline 6.63\% \newline 14.2\% \newline \textbf{3.7e-10} &  266 \newline 11.7\% \newline 49\% \newline \textbf{1.5e-45} &  43 \newline 4.25\% \newline 8.41\% \newline 0.016 &  543 \newline 100\% \newline 100\% \newline \textbf{1e-300} \\ 
  \hline
\end{tabular}
\end{table}

Since we find so many significant overlaps, we expect some genes to be present in multiple prediction sets. Such genes may be more likely to be aging-related than those present only in single set. Indeed, when considering all 2,684 genes predicted by any of the five approaches (including Induced),  while 1,877 (${\sim}70\%$) of them are predicted by exactly one of the approaches (i.e., are in exactly one prediction set), a number of them \emph{are} predicted by multiple approaches. Namely, 499 (${\sim}19\%$), 250 (${\sim}9\%$), 51 (${\sim}2\%$) and 7 (${\sim}0.3\%$) of the 2,684 genes are predicted by  exactly two, exactly three, exactly four, and exactly five approaches, respectively (Supplementary Fig. 11). Results are qualitatively similar when considering only those predicted genes that are absent from any of the aging-related ground truth sets, or those predicted by the four NP approaches but not Induced, (Supplementary Fig. S12). 

Additionally, we examine how many genes are exclusively (i.e., uniquely) present only in a given predicted gene set but not in any of the others.  Observing such genes would indicate that the given gene set (i.e., the corresponding dynamic subnetwork/data integration approach) is capturing at least some complementary aging-related information compared to the other prediction sets. We observe between $16$ and ${\sim}1,300$ such genes, depending on the predicted gene set.  While one could argue that these unique genes could be the result of a statistical bias, for each of the five prediction sets, a considerable number ($>$12\%) of the unique genes are also present in the ground truth data (Supplementary Figs. S13 and S14), which increases our confidence in them.

\subsection{Validation of predicted aging-related genes }
\label{results:validation}

Next, we aim to answer our key question: whether the aging-related gene predictions by at least one of the NP approaches are better, i.e., of higher quality, than the predictions of Induced. To answer this, we measure  overlaps of genes as well as functions (GO terms) between each considered predicted gene set and each aging-related ground truth dataset. Then, first, we quantify the statistical significance of the overlaps  as described in Section \ref{method:overlap}. If the overlaps are higher and more statistically significant for an NP-based prediction set than for the Induced prediction set, this would yield a positive answer to the above question. Indeed, as we show next, this is exactly what we find. Second,  we express the overlaps between a predicted gene set and an aging-related ground truth dataset using precision, recall, and F-score measures. Namely,  precision is the fraction of the predicted genes that are present in the ground truth dataset, recall is the fraction of the genes from the ground truth dataset that are present in the prediction set, and F-score is the harmonic mean of precision and recall. Then, if precision, recall, and F-score are higher for an NP-based prediction set than for the Induced prediction set, this would yield a positive answer to the above question. Indeed, as we show next, this is exactly what we find.

Note that for the above analyses, in addition to the five  sets of predicted genes corresponding to the five dynamic subnetworks/data integration approaches  (Section \ref{results:localcomparison}), here, we define and consider an additional prediction set called \emph{NP-union} that contains all of the predicted genes present in any of the four NP-based prediction sets. We do so in order to analyze the quality of all NP-based predictions as a whole. NP-union contains ${\sim}27\%$ of all 8,938 proteins in the static PPI network. Hence, we now have six prediction sets. Also, for this analysis, we use all three considered ground truth aging-related datasets, i.e., GenAge, GTEx-down and GTEX-up (Section \ref{sect:methods-gtdata}). 

\vspace{0.1cm}

\noindent\textit{\textbf{Gene overlap. }}We measure the overlap of each of the six predicted gene sets with each of the three ground truth sets. A significant overlap with  GenAge, as well as high precision, recall, and F-score with respect to GenAge, would be encouraging, because GenAge is considered to be the most trustworthy source of human aging-related knowledge \cite{De2009}. Even though GTEx-down and GTEx-up come from the same study \cite{jia2018}, they show very different characteristics: GTEx-down is critical for PPIs while  GTEx-up is not (Section \ref{sect:methods-gtdata}). Because of this, and because we use PPI data to predict aging-related genes, we expect a good predicted gene set to overlap more with GTEx-down than with GTEx-up, as well as to have a higher precision, recall, and F-score for GTEx-down than for GTEx-up. In fact, lack of overlap with GTEx-up, as well as low precision, recall, and F-score for GTEx-up, can be interpreted as passing a negative control check.

\begin{figure}[!h]
	\centering
	\includegraphics[scale=0.75]{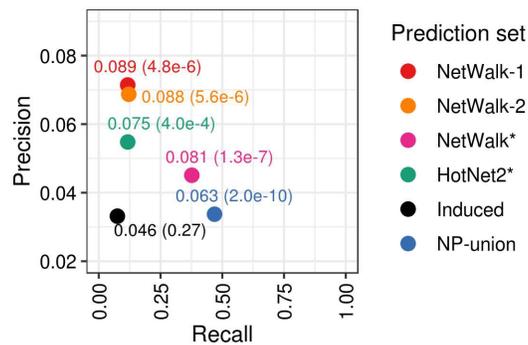}
    \caption{Accuracy (in terms of precision, recall, and F-score) of the six prediction sets with respect to GenAge ground truth data. In the figure, a given point (corresponding to the given prediction set) is labeled with two numbers, where the first number is the F-score of the corresponding prediction set and the second number (in brackets) is the $q$-value of the overlap between the prediction set and GenAge. Analogous results for the other two ground truth datasets, GTEX-down and GTEx-up, are shown in Supplementary Figs. S15 and S16, respectively.}
    \label{fig:prf}
\end{figure}

Below, we discuss the results (Table \ref{tab:tab1}, Fig. \ref{fig:prf}, and Supplementary Figs. S15-S19) first from a ground truth dataset-focused angle and then from a prediction set-focused angle. 

\begin{table}[ht]
\scriptsize
\setlength\tabcolsep{1.5pt}
\centering
\caption{Gene overlaps between the predicted gene sets (columns) and the aging-related ground truth datasets (rows). The numbers in the  table  can be interpreted just as those in Table \ref{tab:tab0}. The table cells with significant $q$-values that improve upon Induced are highlighted in gray. } 
\label{tab:tab1}
\begin{tabular}{|p{1cm}|p{1.2cm}|p{1.2cm}|p{1.1cm}|p{1.1cm}|p{1cm}|p{1.2cm}|}
  \hline
 & NetWalk-1 \newline (396)& NetWalk-2 \newline (422)& NetWalk* \newline (1996)& HotNet2* \newline (511)& Induced \newline (543) & NP-union \newline (2442)\\ 
  \hline
GenAge \newline (239)&  28 \newline 4.6\% \newline 12\% \newline \cellcolor{lightgray!25}{\textbf{4.8e-6}} &  29 \newline 4.6\% \newline 12\% \newline \cellcolor{lightgray!25}{\textbf{5.6e-6}} &  90 \newline 4.2\% \newline 38\% \newline \cellcolor{lightgray!25}{\textbf{1.3e-7}} &  28 \newline 3.9\% \newline 12\% \newline \cellcolor{lightgray!25}{\textbf{4e-4}} &  18 \newline 2.4\% \newline 7.5\% \newline 0.27 &  112 \newline 4.4\% \newline 47\% \newline \cellcolor{lightgray!25}{\textbf{2e-10}} \\ 
   \hline GTEx-down \newline (469) &  43 \newline 5.3\% \newline 11\% \newline \cellcolor{lightgray!25}{\textbf{5.6e-6}} &  47 \newline 5.6\% \newline 11\% \newline \cellcolor{lightgray!25}{\textbf{1.9e-6}} &  176 \newline 7.7\% \newline 38\% \newline \cellcolor{lightgray!25}{\textbf{6.4e-14}} &  33 \newline 3.5\% \newline 7\% \newline 0.176 &  43 \newline 4.4\% \newline 9.2\% \newline \textbf{6.7e-3} &  214 \newline 7.9\% \newline 46\% \newline \cellcolor{lightgray!25}{\textbf{1.7e-17}} \\ 
   \hline GTEx-up \newline (374)& 18 \newline 2.4\% \newline 4.8\% \newline 0.471 &  12 \newline 1.5\% \newline 3.2\% \newline 0.969 &  69 \newline 3\% \newline 18\% \newline 0.969 &  31 \newline 3.6\% \newline 8.3\% \newline 0.035 &  23 \newline 2.6\% \newline 6.1\% \newline 0.594 & 94 \newline 3.5\% \newline 25\% \newline 0.934 \\ 
   \hline
\end{tabular}
\end{table}

\emph{From a ground truth dataset-focused angle}, GenAge significantly overlaps with all of the prediction sets except with Induced. Since GenAge is our most confident set of aging-related ground truth genes and since it shows significant overlap with only the NP-based prediction sets, this is the first and major evidence that NP-based aging-related dynamic subnetworks are more meaningful than the Induced aging-related dynamic subnetwork. GTEx-down significantly overlaps with all of the prediction sets except with that of HotNet2*. GTEx-up does not show significant overlap with any of the prediction sets, as expected.

\emph{From a prediction set-focused angle}, all of NetWalk-1, NetWalk-2, NetWalk*, and NP-union significantly overlap with  GenAge and GTEx-down but not with GTEx-up. HotNet2* shows significant overlap with only GenAge. Induced shows a significant overlap with only  GTEx-down. Hence, at least one, and in fact more than one, of the NP-based prediction sets (NetWalk-1, NetWalk-2, NetWalk*, and NP-union) overlap with more of the ground truth datasets than the Induced prediction set. So, this further confirms that NP-based aging-related dynamic subnetworks are more meaningful than the Induced aging-related dynamic subnetwork.

An additional confirmation is that not only do NetWalk-1, NetWalk-2, NetWalk*, and NP-union significantly overlap with GenAge, while Induced does not, but even for the GTEx-down  that Induced does significantly overlap with, all of the NetWalk-1's, NetWalk-2's, NetWalk*'s, and NP-union's overlaps with  GTEx-down  are higher (more significant) than the Induced's overlap with GTEx-down  (Table \ref{tab:tab1}).  

The above result discussion has focused on the statistical significance of the overlaps. The same conclusions hold in terms of precision, recall, and F-score: multiple NP-based prediction sets are superior to Induced in terms of all three accuracy measures with respect to GenAge (Fig. \ref{fig:prf}) as well as GTEx-down (Supplementary Fig. S15), but not with respect to the negative control-like GTEx-up (Supplementary Fig. S16), confirming superiority of the NP approaches over Induced.

\vspace{0.1cm}

\noindent\textit{\textbf{GO term overlap. }}
We find that each of  the predicted gene sets is significantly enriched in a number of GO terms, between 10 and 127 of them, depending on the prediction set (Table \ref{tab:tab2}). 

\begin{table}[ht]
\scriptsize
\centering
\setlength\tabcolsep{1.5pt}
\caption{GO term overlaps between the predicted gene sets and the aging-related ground truth datasets. The table can be interpreted just as Table \ref{tab:tab1}. 
} 
\label{tab:tab2}
\begin{tabular}{|p{1cm}|p{1.2cm}|p{1.2cm}|p{1.1cm}|p{1.1cm}|p{1cm}|p{1.2cm}|}
  \hline
 & NetWalk-1 \newline (73) & NetWalk-2 \newline (85) & NetWalk* \newline (95)& HotNet2* \newline (10) & Induced \newline (29) & NP-union \newline (127) \\ 
  \hline
GenAge \newline(686) &  33 \newline 4.5\% \newline 45\% \newline \cellcolor{lightgray!25}{\textbf{2.1e-17}} &  34 \newline 4.6\% \newline 40\% \newline \cellcolor{lightgray!25}{\textbf{4.5e-16}} &  39 \newline 5.3\% \newline 41\% \newline \cellcolor{lightgray!25}{\textbf{1.6e-18}} &  10 \newline 1.5\% \newline 100\% \newline \cellcolor{lightgray!25}{\textbf{1.7e-11}} &  12 \newline 1.7\% \newline 41\% \newline \textbf{1.3e-6} &  56 \newline 7.4\% \newline 44\% \newline \cellcolor{lightgray!25}{\textbf{4.2e-28}} \\ 
   \hline GTEx-down \newline (66)&  19 \newline 16\% \newline 29\% \newline \cellcolor{lightgray!25}{\textbf{1.9e-24}} & 15 \newline 11\% \newline 23\% \newline \cellcolor{lightgray!25}{\textbf{1e-16}} &  24 \newline 18\% \newline 36\% \newline \cellcolor{lightgray!25}{\textbf{1.7e-30}} &  4 \newline 5.6\% \newline 40\% \newline \textbf{1e-6} &  11 \newline 13\% \newline 38\% \newline \textbf{1.5e-16} &  32 \newline 20\% \newline 48\% \newline \cellcolor{lightgray!25}{\textbf{2.9e-41}} \\ 
   \hline GTEx-up \newline (15) & 1 \newline 1.1\% \newline 6.7\% \newline 0.135 &  1 \newline 1\% \newline 6.7\% \newline 0.15 &  2 \newline 1.9\% \newline 13\% \newline 0.015 &  0 \newline 0\% \newline 0\% \newline 1 & 2 \newline 4.8\% \newline 13\% \newline \textbf{1.5e-3} & 2 \newline 1.4\% \newline 13\% \newline 0.024 \\ 
   \hline
\end{tabular}
\end{table}

When we measure the overlap between GO terms that are enriched in a given predicted gene set and GO terms that are enriched in a given ground truth dataset, we find that all of the five NP-based prediction sets significantly overlap with GenAge, and all five of them overlaps with GenAge better (more significantly) than Induced (Table \ref{tab:tab2}). Similarly, all five of the NP-based prediction sets  significantly overlap with GTEx-down, and four of them (NetWalk-1, NetWalk-2, NetWalk*, and NP-union) overlap with GTEx-down better than Induced. None of the NP-based predictions significantly overlap with GTEx-up, while Induced does. All of the above results further demonstrate that using NP to create dynamic subnetworks and produce aging-related gene predictions improves upon using Induced to do so.

\vspace{0.1cm}

\noindent\textit{\textbf{Summary. }} Identifying at least one NP-based dynamic  subnetwork that consistently improves upon the Induced dynamic  subnetwork would confirm our hypothesis that NP improves upon the induced approach. In fact, three of the four considered NP-based dynamic  subnetworks, namely NetWalk-1,  NetWalk-2,  and  NetWalk* do so. The remaining NP-based dynamic subnetwork, HotNet2*, does not always improve upon Induced. So, it seems that the NP-based NetWalk approach performs better than not just Induced but also the NP-based HotNet2 approach in our aging-related application. This is somewhat surprising, because HotNet2 is a newer approach that NetWalk, because we gave HotNet2 (and NetWalk) the best-case advantage in terms of method parameter choices,  and because the latter has received significant popularity in the literature as a powerful approach for inference of condition-specific subnetworks. However,  HotNet2 was originally proposed and has typically been used in the context of cancer, which is why it might not generalize well to the context of aging. Nonetheless, NetWalk was neither originally proposed in the context of aging. Yet, it performs quite well in this application.

\subsection{Our NP-based gene predictions are non-random}
\label{sec:validate-net}
If our actual NP-based dynamic  subnetworks and their respective predictions are meaningful, then when we randomize a given dynamic subnetwork and use our actual methodology to make predictions from the randomized data, the number of predictions should be much lower than the number of predictions made from the actual data.

To evaluate whether this is the case, given an actual dynamic subnetwork with 37 age-specific PPI subnetworks, we randomly rewire the edges within each of its 37 subnetworks and then use this collection of 37 randomized subnetworks as a randomized dynamic subnetwork. We keep the age-order of the 37 randomized subnetworks the same (i.e., from younger to older age) as in the actual dynamic network. We perform the randomization procedure 100 times, resulting in 100 randomized dynamic subnetworks for each of the considered actual dynamic subnetworks. Then, we average the results (i.e., the resulting numbers of predictions) over the 100 random runs. Because  NetWalk-1, NetWalk-2, and NetWalk* are the three dynamic subnetworks that consistently improve upon Induced while HotNet2* does not (see above), in this analysis, we consider only the former three dynamic subnetworks. 

Indeed, we find that for each of  NetWalk-1, NetWalk-2, and NetWalk* dynamic subnetworks, the number of predictions is much larger (${\sim}600\%$, ${\sim}800\%$,  and ${\sim}1400\%$, respectively) from the actual data than from the corresponding randomized data (Supplementary Fig. S20). We compute the statistical significance of this result as follows. For each of the dynamic subnetworks, we evaluate the probability of getting from the randomized data the same or higher number of predictions as from the actual data. To do this, we use the concept of $z$-scores. Given the 100 prediction counts from the randomized data, i.e., their average, $z$-score measures how many standard deviations the actual count is from the average \cite{Faisal2014}. The higher the $z$-score, the lower the  corresponding  $p$-value, and less likely it is to obtain the actual count by chance.  Indeed, we observe extremely high $z$-scores $>43$ and extremely low corresponding $p$-values $<10^{-300}$) for the three dynamic subnetworks. Hence, our results are non-random.

\subsection{Literature validation of novel NP-based predicted genes}
\label{sect-results-litval}
We identify novel NP-based predicted genes, i.e., all genes that are present in each of the four NP-based predicted gene sets (i.e., NetWalk-1, NetWalk-2, NetWalk*, and HotNet2*) but are absent from all of the aging-related ground truth datasets as well as from the Induced predicted gene set. There are 16 such genes (Supplementary Fig. S21). We successfully validate nine of them in the literature as being related to either the aging process directly or to a disease that is known to be aging-related: ECH1, MAP2K1, SPARCL1, SKP1, BAG1, SNTA1, STX16, DYNLL1, and SHC3  (Table \ref{tab:lit-validation}).  Since all four NP-based dynamic subnetworks identify these 16 genes as aging-related, and since we validate (using literature search) nine (i.e., majority) of these genes as aging-related, the remaining seven genes can be considered as novel gene predictions that are potentially (if not likely) aging-related. The seven genes are: F8, MED19, MPDZ, NCOA4, PIP4K2A, RAP1GDS1, TF. 

\begin{table}[!h]
\centering
\scriptsize
\caption{List of the nine out of 16 genes predicted by all four NP approaches that we validate in PubMed. We show the PubMed IDs (PMIDs) of the articles that provided the validations.}
\begin{tabular}{|c|p{5cm}|c|}
  \hline
 Gene symbol& Description& PMID \\ 
  \hline
  BAG1 &Its under expression with age has been shown in articular cartilage, showing evidence of the involvement of BAG1 in the regulation of cartilage aging&  15278942 \\
  \hline
    DYNLL1 & Its downregulation has been shown in the late onset of aging-related Alzheimer's disease& 18789830\\
  \hline
  ECH1 & Its orthologous gene \emph{ech1} (found in rat) was shown to be upregulated in the aged basal forebrain cholinergic neurons& 17560690\\
  \hline 
   MAP2K1 & Its orthologous gene \emph{Map2k1} (found in mouse) was shown to be down-regulated in aged heart & 19031007\\
  \hline
  SHC3 &  It is has been shown to be a potential therapeutic target against Alzheimer's disease & 17170108\\ 
\hline 
    SKP1 & It has been shown to be involved in the regulation of Caenorhabditis elegans lifespan& 17392428\\
    \hline 
      SNTA1 & Its down-regulation was shown in old mouse skeletal muscle& 30089464\\ 
   \hline
    SPARCL1 & Its upregulation was shown in aged astrocytes of mouse brain & 29437957\\
  \hline
     STX16 & Its downregulation has been shown in the onset of aging-related Alzheimer's disease, which has been related to aging \cite{Kerchner2016}.&     18572275\\
  \hline

\end{tabular}\label{tab:lit-validation}
\end{table}

\subsection{Our results are robust to the choice of static PPI network data}
\label{sect:results-robust}

Thus far, we have used the static PPI network from HPRD.  To verify that our key result -- NP-based dynamic aging-related subnetworks being more meaningful than the Induced dynamic aging-related subnetwork -- is robust to the choice of static PPI network data, we perform all of our above analyses using an alternative human PPI network, i.e., HINT+HI2 (Section \ref{sec:methods-ppidata}), as follows. 

Given the static HINT+HI2 PPI network, we identify aging-related dynamic subnetworks using the same four versions of the considered NP approaches (i.e., HotNet2, HotNet2*, NetWalk, and NetWalk*) plus Induced, following the same methodology that we have used thus far in the paper (Section \ref{sect:methods}). This results in a dynamic subnetwork for Induced, no dynamic subnetwork for HotNet2 or HotNet2*, a dynamic subnetwork for NetWalk, and two dynamic subnetworks for NetWalk*. We use each of the above four dynamic subnetworks (i.e., one Induced subnetwork and  three NP-based subnetworks) to predict aging-related genes using the same methodology that we have used thus far in the paper. Finally, we examine: 1) overlaps of the four prediction sets with one another, 2) gene overlaps between each of the predictions sets and each of the three ground truth datasets, and 3) GO term overlaps between each of the prediction sets and each of the ground truth dataset. 

Similar to our results when using the HPRD network, we find the following when using the HINT+HI2 network: 1) all of the predicted gene sets significantly overlap with each other; 2) at least one of the NP-based predicted gene sets shows a more significant gene overlap with the ground truth data than the predicted Induced gene set; and 3) at least one of the NP-based predicted gene sets shows a more significant GO term overlap with the ground truth data than the predicted Induced gene set (Supplementary Tables S3-S5). Thus, our results are robust to the choice of static PPI network data.

\section{Conclusion}

We  have hypothesized that using  network propagation to integrate aging-related gene expression data with static PPI network data in order to infer a dynamic aging-related  PPI subnetwork would result in higher-quality aging-related gene predictions from the inferred subnetwork compared to the only existing approach for this purpose, i.e., the induced approach.  Indeed, this is what we have found.  

To be able to evaluate the above hypothesis, we had to methodologically extend the existing network propagation approaches, as they currently allow only for inference of a static but not dynamic condition-specific subnetwork.

As is typically done \cite{Faisal2014, elhesha2019identification}, we create a dynamic aging-related PPI subnetwork such that each of its age-specific subnetworks (i.e., temporal snapshots) corresponds to one of the ages present in the gene expression data. Specifically, given 37 ages present in our considered gene expression data, our dynamic subnetwork is a collection of the 37 corresponding age-specific snapshots.  Alternatively, it may be beneficial to (i) bin the 37 ages into different age groups, by combining adjacent ages that are ``similar enough'' into a single age group, (ii) then form one age-specific PPI subnetwork per age group (rather than per individual age), and (iii) finally combine the age group-specific (rather than individual age-specific) PPI subnetworks into a dynamic, aging-related  subnetwork. However, figuring out a proper ``temporal resolution'', i.e., an appropriate time interval length to be used to construct a temporal snapshot  (in our case, an appropriate age interval length to be used to construct an age group-specific snapshot) is a non-trivial challenge \cite{Meng2016}. For example, should one construct an age group-specific snapshot for each 5 years of the human lifespan, or each 10 or more years? Should all age group-specific snapshots span equal-length age intervals, or should some span longer intervals while others span shorter intervals? A possible solution to determining appropriate values of these parameters may be to computationally identify $k$ time points in the entire time interval where network structure (with respect to some network structural property) significantly changes, and using the $k+1$ temporal segments separated by these changing time points to construct the  $k+1$ corresponding temporal snapshots \cite{hulovatyy2016scout}. However, even in this case, it is unclear exactly which network structural property to rely on. Moreover, even if the above challenge of determining appropriate age groups would be resolved, it would still remain an open question of how to use, i.e., combine, gene expression data for multiple ages in a given group to determine whether two genes should be linked in the snapshot corresponding to that age group. In summary, examining an appropriate way of choosing the dynamic network construction parameters is a complex research question of its own that would warrant a separate research study. Consequently, this is out of the scope of this paper and is the subject of future work.

As a part of our study, we have predicted novel human aging-related genes and validated a majority of them in the literature. Our framework for dynamic condition-specific inference and analysis  introduced in this paper can be applied to studying other dynamic biological processes, such as disease progression over time.


\section*{Acknowledgements}

This work was supported by the National Science Foundation (CAREER CCF-1452795).

\section*{Additional information}

\textbf{Competing financial interests:} The authors declare no competing financial interests.

\clearpage

\end{document}